\documentclass{aa}  

\usepackage{graphicx}
\usepackage{txfonts}
\usepackage{lipsum}
\usepackage{subcaption}         
\usepackage{lscape}             
\usepackage{placeins}           
                                
\usepackage[colorlinks=true,linkcolor=blue,urlcolor=blue,citecolor=blue]{hyperref}

\begin{document}

   \title{Dust evolution during protoplanetary disk buildup enhances CO~ice relative to water}



   \author{J. Dr\k{a}\.{z}kowska
        }

   \institute{Max Planck Institute for Solar System Research, Justus-von-Liebig-Weg 3, 37077 G\"ottingen, Germany\\
             \email{drazkowska@mps.mpg.de}}

   \date{Received January 27, 2026; accepted March 25, 2026 }

 
  \abstract
   {Water ice is expected to be the dominant volatile component of bodies formed in the outer Solar System. However, recent observations of comets and trans-Neptunian objects suggest that the relative abundances of ices can vary substantially, with some bodies exhibiting unusually high CO/H$_2$O ratios.}
   {We study the prospects of producing CO-rich pebbles and planetesimals.} 
   {We use a one-dimensional protoplanetary disk model with dust evolution including coagulation, fragmentation, and radial drift, water and CO ice and vapors evolution, and planetesimal formation via the streaming instability. We compare models with and without the disk formation stage.}
   {CO-rich pebbles can be formed at the CO snow line due to the cold finger effect, regardless of whether the disk buildup is included. Models including disk buildup show stronger CO enhancement relative to water in the outer disk. However, CO-rich planetesimals do not form in the smooth disk models.}
   {The formation of CO-rich planetesimals likely requires mechanisms that preserve the CO-enriched ice reservoir, such as pressure traps or gas removal processes. Models concerning the chemical evolution of protoplanetary disks and its impact on the atmospheric C/O ratio of forming planets should consider the disk buildup stage.}


   \maketitle
   \nolinenumbers

\section{Introduction}

Water is nominally the most abundant volatile in the Solar System. Comets forming in the cold outer region of the Solar nebula are expected to be a mix of refractory and icy dust species, with water ice dominating the mixture with the abundance of over 50\% \citep{Lodders2003}. However, some of the known comets seem to be dominated by CO ice rather than water ice. C/2009 P1 (Garradd) was the first comet for which a significant CO/H$_2$O production rate was observed inside the water snow line \citep{Feaga2014}. Although it did not cross inside the water snow line, C/2006 W3 (Christensen) is also suspected to have a high CO abundance based on its production rate \citep{Bonev2017}. C/2016 R2 (PanSTARRS) is the canonical example of a comet with CO dominating the volatile budget with an estimated CO/H$_2$O ratio as high as 300 \citep{Biver2018, McKay2019, Harrington2022}. The interstellar comet 2I/Borisov showed unusually high CO compared to typical Solar System comets \citep{Cordiner2020, Bodewits2020}. The recent JWST observations revealed distinct surface compositions of the trans-Neptunian objects, with water ice not universally present, suggesting their compositional diversity \citep{PinillaAlonso2025, DePra2025}.

The meteoritic evidence also suggests that water was not always the dominant ice component in the carbonaceous chondrite parent bodies that formed beyond the water ice line. The initial water-to-rock ratio varied from 0.01 to above 1 in different carbonaceous chondrite classes \citep{Brearley2006, Marrocchi2018}. \citet{Sridhar2021} showed that some of the carbonaceous chondrite parent bodies, in particular Renazzo-type (CR) and the ungrouped Tagish Lake (TL) chondrites, must have included a significant fraction of ammonia ice as they were altered by the action of an ammonia-bearing fluid based on the distinct magnetite morphologies corresponding to different aqueous alteration pathways found in those meteorites.
 
Clathration is often discussed as a possibility to locally concentrate CO and other volatiles in comets. In the protoplanetary disk, CO can become trapped in the water ice more efficiently than expected from equilibrium condensation alone \citep{BarNun1985, Mousis2010}. When a cometary nucleus made of such clathrates warms up, the trapped CO is released preferentially, increasing the observed CO/H$_2$O ratio. However, \citet{Mousis2021} showed that the amount of water required by clathration to match the N$_2$/CO ratio measured for C/2016 R2 (PanSTARRS) is inconsistent with its estimated low abundance.

\citet{Price2021} and \citet{Mousis2021} proposed that enhancement of CO is expected locally in the vicinity of the CO evaporation front due to the so-called cold-finger effect: the recondensation of CO vapor diffusing outwards across the CO condensation line. However, these models start the ice and vapor evolution from a fully-fledged disk, neglecting the disk buildup stage. In this paper, we revisit the expected CO enhancement using a model including the protoplanetary disk formation stage and compare the results to a model starting with a fully-fledged disk.

This paper is organized as follows. We present the numerical setup in Sect.~\ref{sect:methods}. We present the results of our models in Sect.~\ref{sect:results}. We discuss the limitations of the numerical approach and implications of the results in Sect.~\ref{sect:discussion}. Section~\ref{sect:conclusions} concludes this work.

\section{Methods}\label{sect:methods}

We build a one-dimensional model of a protoplanetary disk following the methodology presented in \citet{DD2018} (hereafter DD18). The protoplanetary disk formation is considered by the infall of the spherically symmetric rotating molecular cloud \citep{Shu1977, Ulrich1976, Hueso2005}. We consider a molecular cloud of one Solar mass rotating with the rate of $\Omega=7\cdot10^{-15}$~s$^{-1}$, which ends up forming a single star and the surrounding protoplanetary disk, which reaches its peak mass of 0.25~M$_\odot$ after 0.64 Myrs since the beginning of the infall simulation, which we will be calling time zero when analyzing the results. The infall proceeds inside-out such that initially the material accretes directly onto the star and the centrifugal radius crosses the inner boundary of the disk about 0.14 Myrs after the start of the model, corresponding to time $-0.5$~Myrs marked in the figures. We assume that the disk has an intrinsic turbulence prescribed with $\alpha=10^{-3}$ and that its temperature structure is set both by the viscous heating and irradiation from the central star.

We run sets of two models, one initializing dust at a constant dust-to-gas ratio across the disk when the gas disk is at its maximum mass (models without disk buildup) and a corresponding model where dust evolution was considered from the very beginning, when the first material arrived at the disk (models with disk buildup). The dust has three components: refractory dust with 0.5\% mass ratio, water ice with 0.5\% mass ratio, and CO ice with 0.14\% mass ratio with respect to the gas. We assume that dust aggregates are spherical and that there is one typical grain size $a_{\rm{p}}$ at each radial distance, which is calculated according to the same procedure as in DD18, including initial growth phase, fragmentation, and radial drift, similar to the two population model proposed by \citet{Birnstiel2012}. We assume that the fragmentation speed is 10~m~s$^{-1}$ outside and 1~m~s$^{-1}$ inside of the water snow line with a smooth transition in between. The dust surface density of each dust component is evolved by radial drift, advection with the gas flow, and diffusion. As in DD18, the water ice can evaporate to water vapor, which can recondense back to water ice.

We extended the model presented by DD18 to include the CO ice and vapor evolution. In the same way as for water, we calculate the equilibrium pressure of CO at every location of the disk
\begin{equation}
    P_{\rm{eq}}^{\rm{CO}} = P_{\rm{eq,0}}^{\rm{CO}} {\mathrm e}^{-T_a^{\rm{CO}}/T},
\end{equation}
where we adopted $P_{\rm{eq,0}}^{\rm{CO}}= 7.73 \cdot 10^{11}$~g~cm$^{-1}$~s$^{-2}$ and $T_a^{\rm{CO}}=1030$~K following \citet{Leger1985}, consistent with later laboratory determinations. 

We follow the evolution of the surface density of CO ice $\Sigma_{\rm{ice}}^{\rm{CO}}$ which changes together with the other dust species by the radial drift, advection with gas flow, and diffusion, and CO vapor $\Sigma_{\rm{vap}}^{\rm{CO}}$ which is evolved according the the accretion flow of gas and diffusion. Adopting the approach that \citet{Schoonenberg2017} proposed for water, $\Sigma_{\rm{ice}}^{\rm{CO}}$ and $\Sigma_{\rm{vap}}^{\rm{CO}}$ also change due to CO condensation and evaporation, and the rate of this change is given by
\begin{equation}
   \dot{\Sigma}_{\rm{ice}}^{\rm{CO}} = -\dot{\Sigma}_{\rm{vap}}^{\rm{CO}} = R_{\rm{c}}{\Sigma}_{\rm{ice}}^{\rm{CO}} \Sigma_{\rm{vap}}^{\rm{CO}} - R_{\rm{e}}{\Sigma}_{\rm{ice}}^{\rm{CO}}, 
\end{equation}
where the condensation and evaporation rates are defined as
\begin{equation}
    R_{\rm{c}} = 6 \sqrt{\frac{k_{\rm{B}}T}{\mu_{\rm{CO}}}}\frac{1}{\pi a_{\rm{p}} \rho_{\rm{p}} H_{\rm{g}}}
    \label{eq:Rcond}
\end{equation}
and
\begin{equation}
    R_{\rm{e}} = 6 \sqrt{\frac{2\pi\mu_{\rm{CO}}}{k_{\rm{B}}T}{}}\frac{P_{\rm{eq}}^{\rm{CO}}}{\pi a_{\rm{p}} \rho_{\rm{p}}},
    \label{eq:Revap}
\end{equation}
where $k_{\rm{B}}$ is the Boltzmann constant, $T$ is the midplane disk temperature, $\mu_{\rm{CO}} = 28 m_{\rm{H}}$, with $m_{\rm{H}}$ being the proton mass, is the weight of CO molecule, $a_{\rm{p}}$ and  $\rho_{\rm{p}}$ are dust grain radius and material density, and $H_{\rm{g}}$ is disk scale-height. It is worth noting that both the condensation and evaporation rates increase when dust grains are smaller.

We use 1000 logarithmically spaced radial cells to cover the radial grid stretching from 0.3~au to 10~pc. We run the model until 3~Myr after the time the disk reached its peak mass (start of the model without disk buildup). The numerical code used to produce results presented in this paper is available at \url{https://github.com/astrojoanna/DD-diskevol-CO}.

\section{Results}\label{sect:results}

\subsection{Fiducial model}
    
   \begin{figure*}
        \centering
        \includegraphics[width=0.85\textwidth]{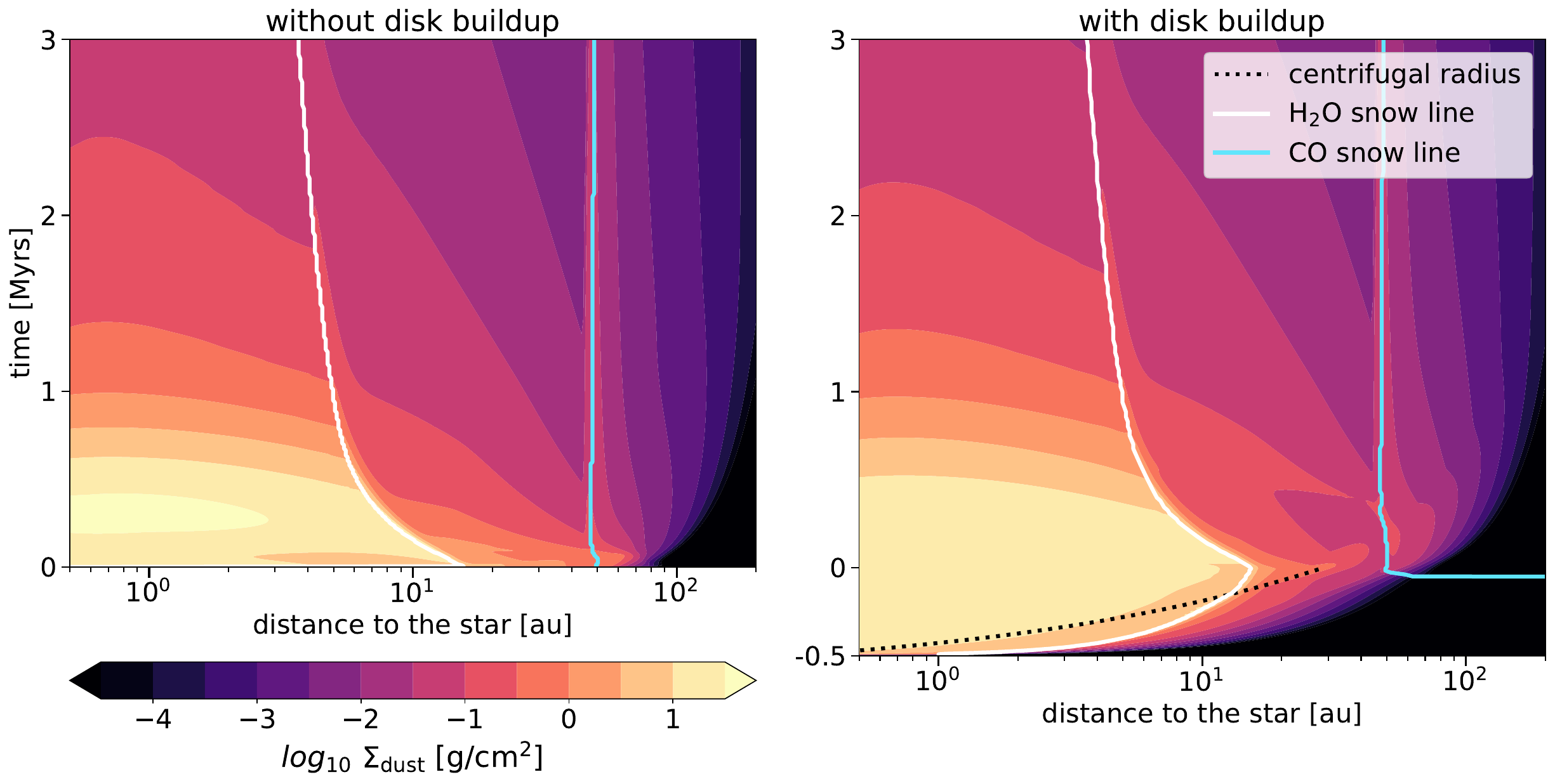}
        \caption{Dust surface density evolution in the fiducial model without disk buildup (left) and with disk buildup taken into account (right). Time zero refers to the moment the gas disk reaches maximum mass in the model including its buildup, corresponding to the time the dust is initialized in the model without disk buildup.}
        \label{fig:sigmad}
    \end{figure*}

\begin{figure*}
        \centering
        \includegraphics[width=0.85\textwidth]{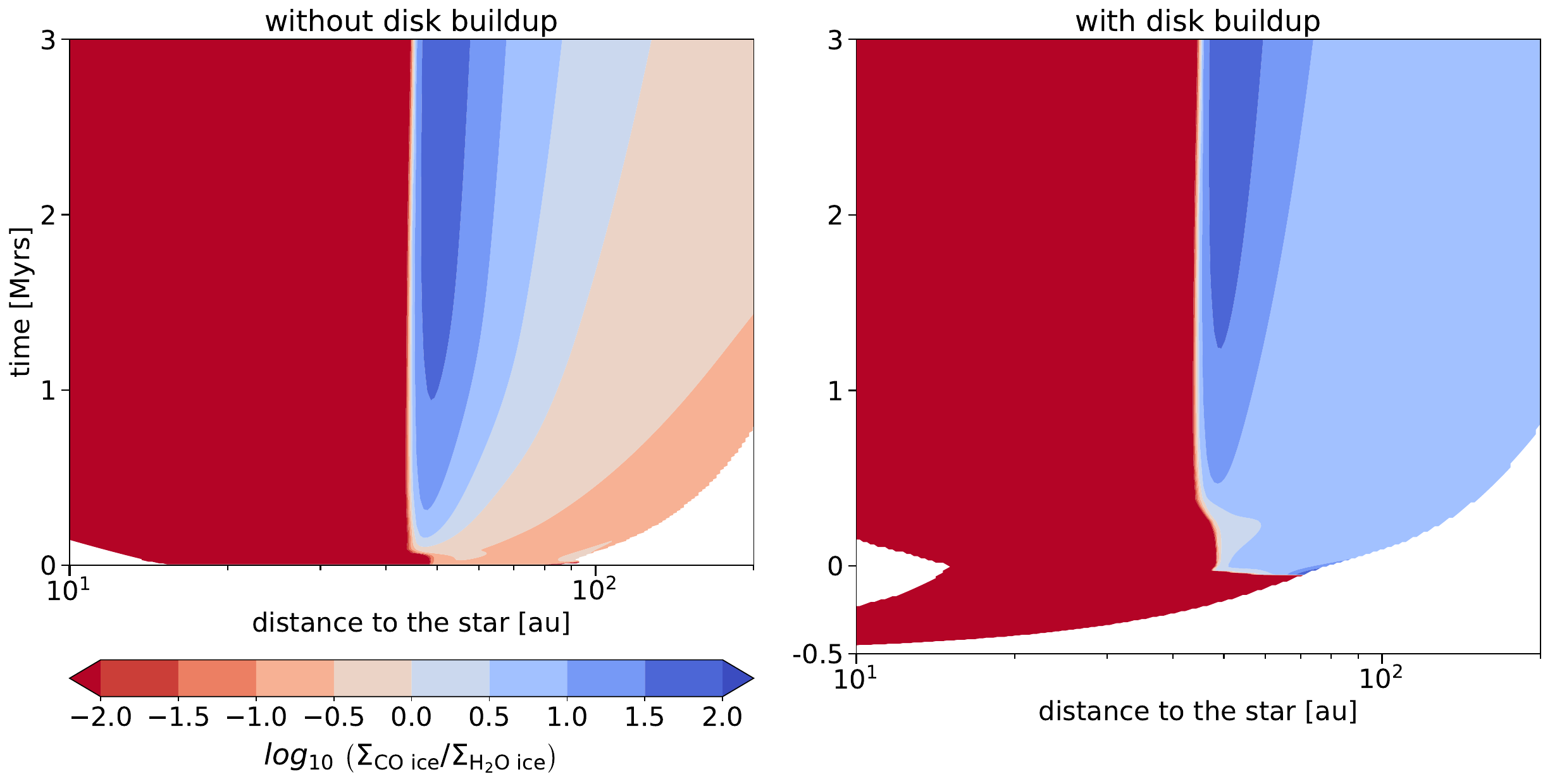}
        \caption{Evolution of the CO to H$_2$O ice ratio as a function of time and space in the fiducial model without disk buildup (left) and with disk buildup taken into account (right). Note the x-axis is different from Fig.~\ref{fig:sigmad} as this figure is centered at the CO snow line.}
        \label{fig:COtoH2O}
\end{figure*}

Figure~\ref{fig:sigmad} shows the evolution of the total dust surface density, that is, the sum of refractory dust and ice components, in the run without and with dust evolution during buildup taken into account. Both models evolve in a similar way, with the main difference being slightly faster inner disk depletion in the model including disk buildup. This is caused by the dust evolution during the buildup phase lasting over 0.5 Myrs. Because the dust already had time to grow and drift inwards, there is more dust in the inner disk, and the outer disk is already depleted compared to the run in which the dust is only initialized at the time of maximum gas disk mass (marked as time zero in the plots). This leads to the inner disk being depleted in dust faster than in the run without disk buildup, consistent with the conclusion reached in \citet{Birnstiel2010}.

We define the location of the water and CO snow lines as the radial distance where 50\% of the total surface density of the respective species is in the gas phase and 50\% in the solid phase. The position of the water and CO snow lines is practically identical in both models as they share the same temperature structure, which is independent of dust surface density and maximum grain size. The water snow line is located in the part of the disk where temperature is set by viscous heating, and it moves inwards with time as the disk becomes less massive and cools down. On the contrary, the position of the CO snow line does not change significantly during the disk evolution, as it is located in the irradiation-dominated part of the disk, and we do not include stellar luminosity evolution. This is consistent with findings of the earlier work of \citet{Stammler2017}. The local increase in the solids surface density is visible outside of both the snow lines is caused by the cold-finger effect, with additional contribution of the traffic jam effect in the case of the H$_2$O snow line, see \citet{Drazkowska2017} and DD18 for details. In agreement with the results of DD18, we find no planetesimal formation in the models assuming the global, isotropic turbulence with $\alpha=10^{-3}$. 

It is worth noting that in the model with disk buildup, the centrifugal radius shifts from inside to outside of the water snow line during the infall phase. As a result, water delivered early in the buildup is deposited inside the snow line in vapor form and later recondenses, whereas material accreted after the centrifugal radius has moved outward arrives as primordial ice. Presence of primordial water ice seems to be necessary to explain the deuterium-to-hydrogen enrichment of water in the Solar System \citep{Cleeves2014}. However, the situation is different for the CO, as the CO snow line stays outside of the centrifugal radius at all times and thus the primordial CO ice evaporates during the infall phase and recondenses when the CO vapor is carried outside of the CO snow line, in agreement with the findings of the 2D model presented by \citet{Visser2009}. Thus, the abundance of CO ice is solely regulated by its condensation rate, whereas the water ice reservoir reflects both primordial delivery and recondensation. This difference in origin and regulation underlies the distinct behavior of CO and water ice abundances presented below.

\begin{figure*}
        \centering
        \includegraphics[width=\textwidth]{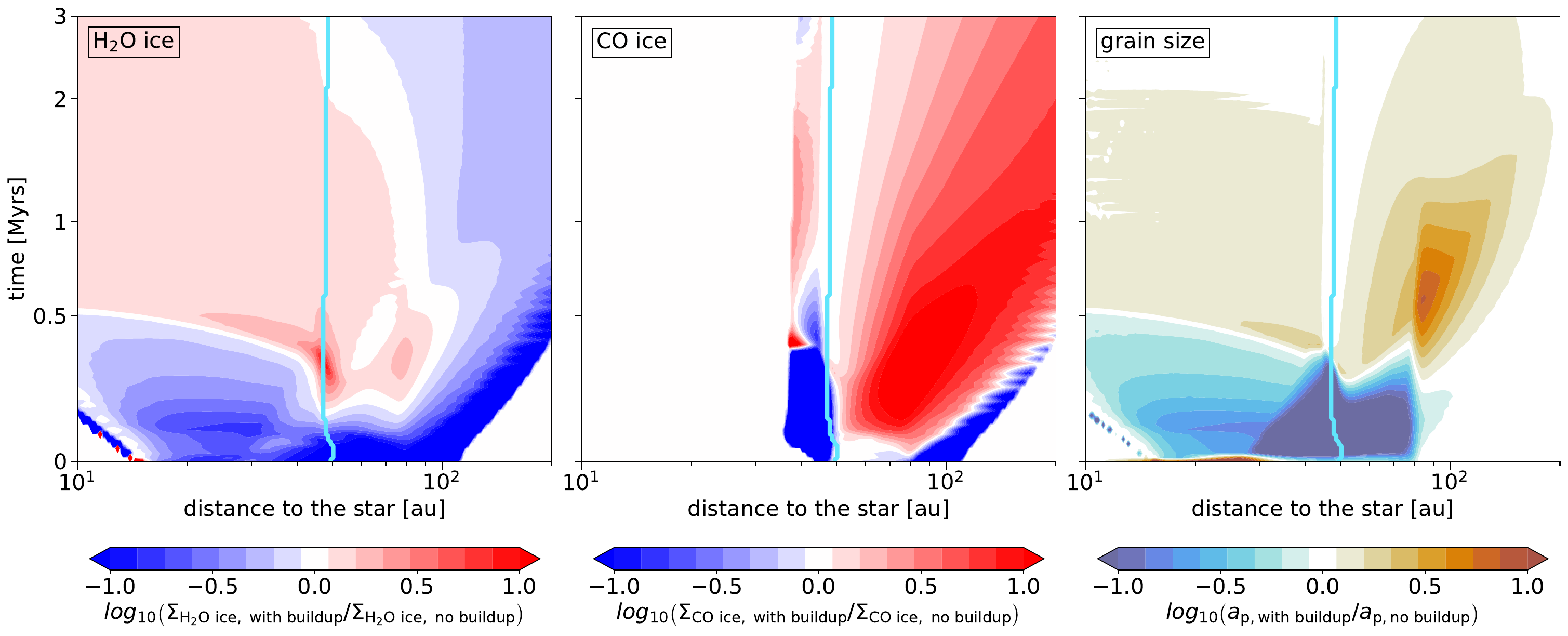}
        \caption{Difference between the evolution of the surface density of H$_2$O ice (left), CO ice (middle) and dust grain size (right) between the runs with and without disk buildup. The vertical blue line marks the location of the CO snow line.}
        \label{fig:difference}
\end{figure*}

\begin{figure}
   \centering
   \includegraphics[width=\hsize]{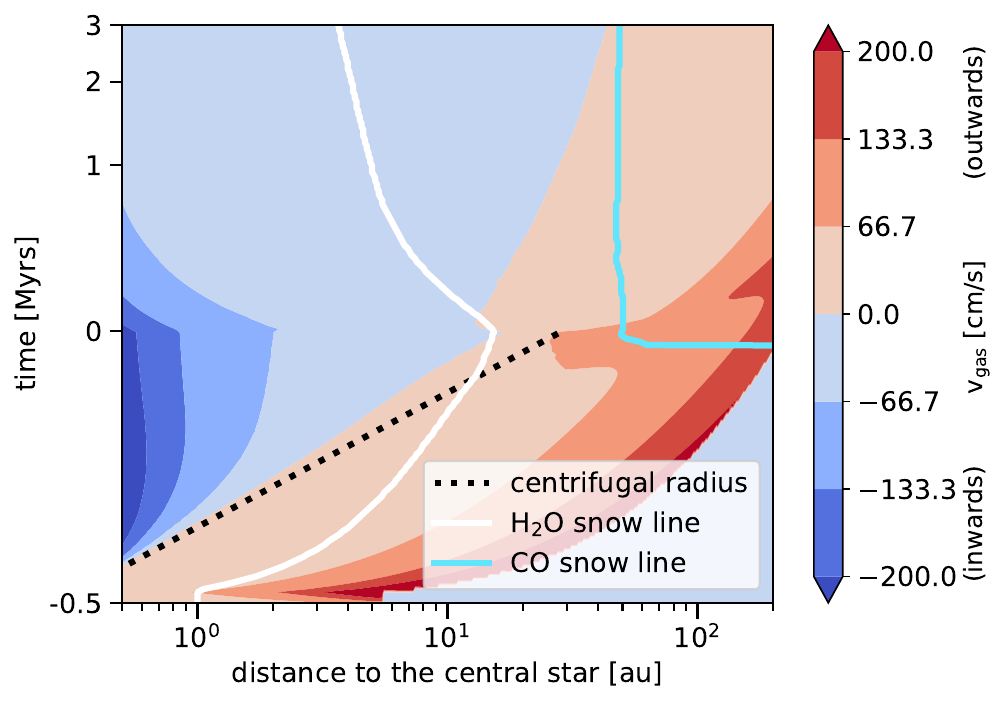}
      \caption{Radial gas velocity in the fiducial model with disk buildup as a function of radial distance and time. Blue colors mean the velocity vector is directed towards and red colors away from the central star.}
         \label{fig:vgas}
\end{figure}

Figure~\ref{fig:COtoH2O} shows the evolution of the ratio of CO ice to water ice surface density in both models. Here, we observe a significant difference between the two runs. In the model without disk buildup, there is a cold-finger-driven enhancement of CO ice outside of the CO snow line exceeding a factor of 10 after 1~Myr of evolution but the CO/H$_2$O ratio gradually decreases with radial distance to the fiducial value of 0.28 (corresponding to the log10 value of -0.55 in the figures), consistent with the earlier results of \citet{Price2021} and \citet{Mousis2021}. However, in the model including disk buildup, the whole region outside of the CO snow line becomes dominated by the CO ice, with the CO/H$_2$O ratio of 3. The reason for this difference is illustrated in Fig.~\ref{fig:difference}. Due to dust evolution during disk buildup, the outer disk is depleted in solids compared to the model in which dust evolution starts when the disk is fully formed. This is reflected in the lower water ice density (left panel of Fig.~\ref{fig:difference}), and the consequence is the smaller dust grain size (right panel of Fig.~\ref{fig:difference}). The dust size is smaller because the lower dust-to-gas ratio leads to slower dust growth and a stronger radial drift barrier \citep[see, e.g.,][]{Birnstiel2012}. However, since the condensation rate given by Eq.~(\ref{eq:Rcond}) is higher for the smaller grain size, a larger fraction of the CO reservoir is transferred from the gas to the solid phase via freeze-out compared to the model without buildup, leading to a higher surface density of CO ice (see the middle panel of Fig.~\ref{fig:difference}). This is promoted by the fact that the centrifugal radius is inside the CO snow line and thus the CO vapor is carried outwards with the gas flow (see Fig.~\ref{fig:vgas}), leading to all CO ice coming from freeze-out. Due to the outward flow of CO vapor, the higher condensation rate is not compensated by evaporation, whose rate is also higher for smaller grains (see Eq.~\ref{eq:Revap}). The higher density of CO ice in the model with buildup ultimately leads to larger grain sizes after 0.5~Myr of evolution (right panel of Fig.~\ref{fig:difference}).

\begin{figure}
   \centering
   \includegraphics[width=\hsize]{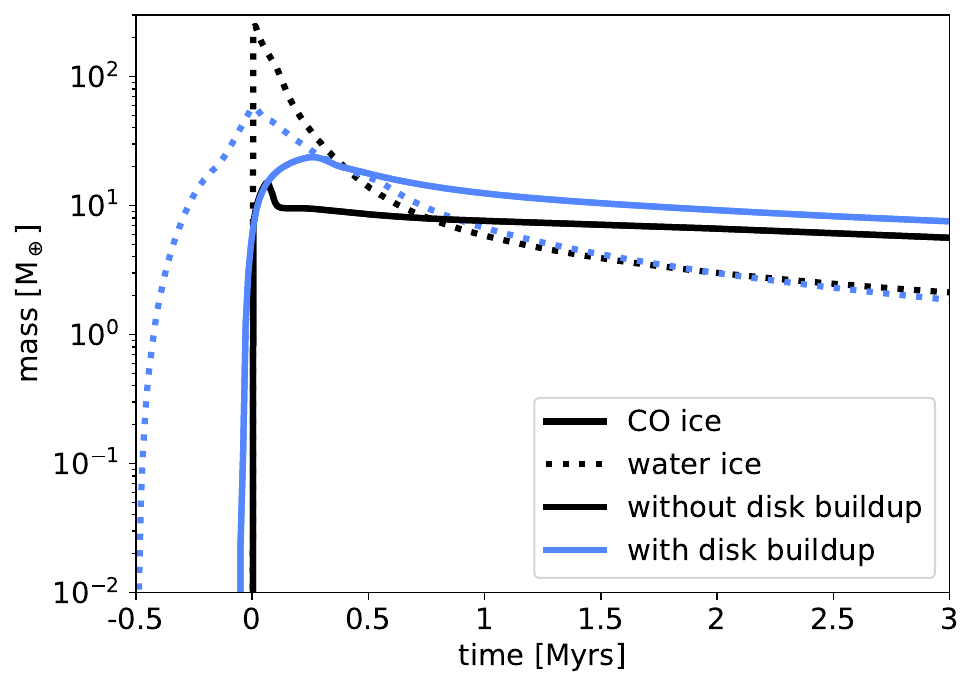}
      \caption{Time evolution of the total mass of CO ice and water ice (as indicated by the different line styles) in the models with and without disk buildup (indicated with blue and black lines, respectively) for the fiducial set of models.}
         \label{fig:massbudget_fid}
\end{figure}

The mass budget analysis shown in Fig.~\ref{fig:massbudget_fid} confirms that the difference between the two models revealed by Fig.~\ref{fig:COtoH2O} is primarily associated with enhanced CO freeze-out shortly after the disk reaches its maximum mass. At $t=0.1$~Myrs, the total mass of CO ice is 14 Earth masses higher in the model with disk buildup and this difference gradually drops to 2 Earth masses at 3~Myrs. It is worth noting that the first CO ice is formed just before the disk reaches its maximum mass in this model. Thus, the CO ice component does not yet have a chance to significantly evolve by radial drift. The situation is different for the water ice, the evolution of which starts earlier because of the inside-out buildup pattern. Over one hundred Earth masses of water ice are already lost from the disk during the buildup phase. The total mass of water ice becomes practically equal in both models after the first quarter million years of evolution, because the surplus water ice present in the model without disk buildup is lost due to faster radial drift caused by faster grain growth. However, the effect of enhanced CO freeze-out on the smaller dust grain sizes caused by the depletion of water ice in the outer disk at the epoch of initial CO ice condensation propagates until the end of the simulation. This is because the outer part of the disk, to which the CO ice is confined, is driven not only by the radial drift but also by the viscous spreading of the disk. As shown in Fig.~\ref{fig:vgas}, during all of the evolution, the CO snow line is located outside of the critical radius, so the gas and CO vapor are flowing outwards, enhancing the cold finger effect. This slows down the loss of CO ice compared to the water ice and allows the initial difference to propagate until the end of our model. Our results demonstrate that the CO/H$_2$O ratio in solids depends not only on the final disk structure, but on the evolutionary path by which the disk formed.

\subsection{Lower dust diffusivity and planetesimal formation}

As mentioned in the previous subsection, in the fiducial set of models, no planetesimals are formed. This is consistent with the findings of DD18 showing that planetesimal formation is possible if the dust diffusivity is reduced relative to that of the gas. This corresponds to a situation in which the disk midplane, where the dust is concentrated, is less turbulent than the disk as a whole. It is a typical assumption made in models dealing with planetesimal formation \citep[see, e.g.,][]{Carrera2017, Ercolano2017}. To test how the planetesimal formation changes the CO ice enhancement, we performed models with reduced dust diffusivity prescribed as $\alpha_{\rm{dust}}=10^{-5}$, corresponding to the model highlighted in DD18.

\begin{figure*}
        \centering
        \includegraphics[width=0.85\textwidth]{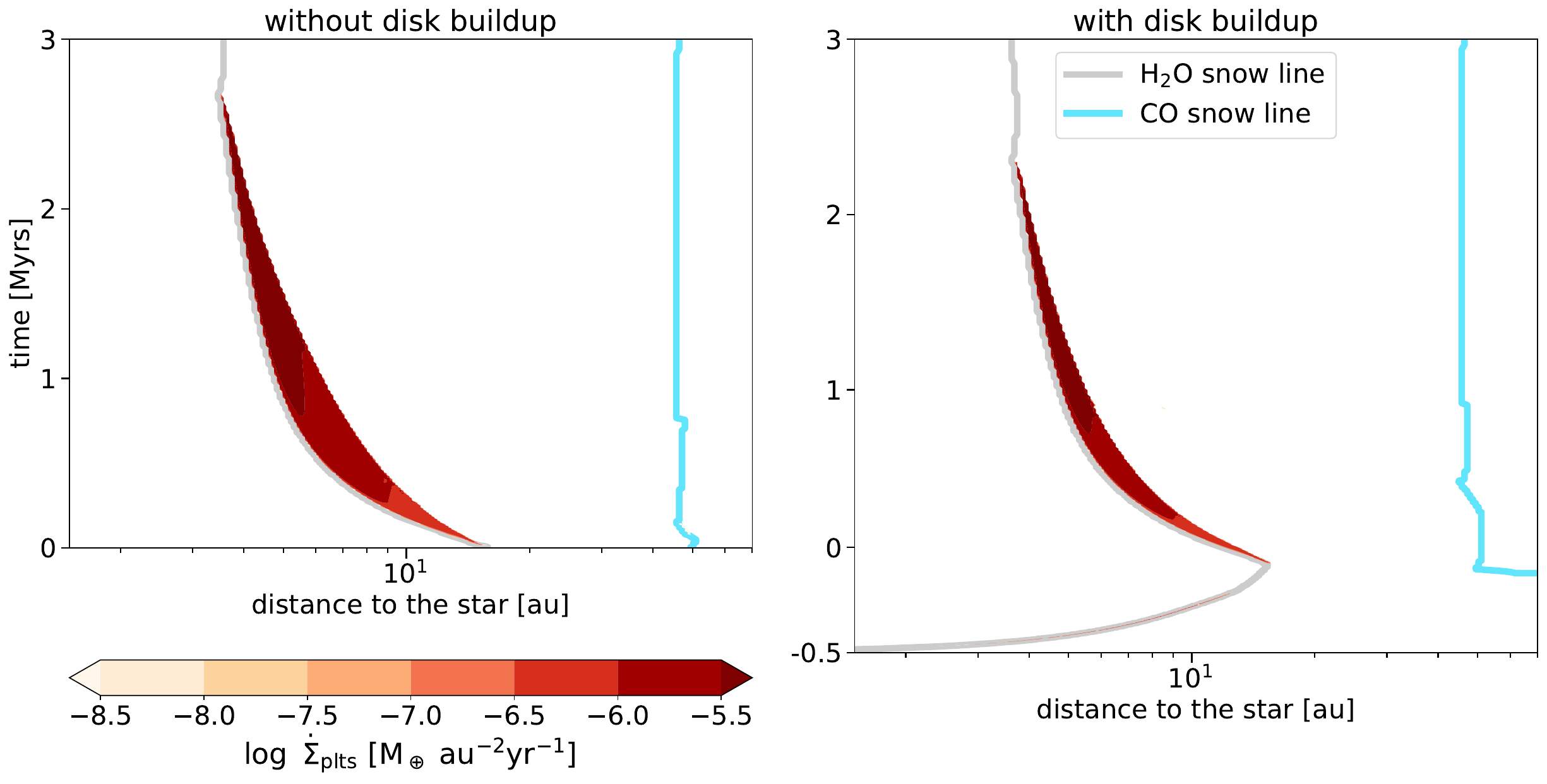}
        \caption{Planetesimal formation rate as a function of time and space in the models with lower dust diffusivity. The left panel corresponds to the model without disk buildup and the right panel to the model with disk buildup taken into account.}
        \label{fig:planetesimals}
\end{figure*}

Figure~\ref{fig:planetesimals} shows the planetesimal formation rate in the models without and with disk buildup included. As in DD18, planetesimal formation is only possible outside of the water snow line. The dust-to-gas ratio does not become high enough at the CO snow line to reach the planetesimal formation threshold, which we set to unity. In the model starting with a fully-fledged disk, the midplane dust-to-gas ratio reaches 0.9 after 60 000 years of evolution and stays at this level for about 50 000 years, after which it drops and stabilizes at around 0.05. This could grant some planetesimal formation if we used more favorable criteria, that is, a lower midplane dust-to-gas ratio threshold. For example, \citet{Gole2020} suggested the threshold midplane dust-to-gas ratio of 0.5 based on a set of hydrodynamic models with external turbulence included and \citet{Carrera2025} suggested the threshold value may be as low as 0.3 if a feedback between dust coagulation and streaming instability is taken into account. In any case, planetesimal formation at the CO snow line would operate only for a limited time and only in the model without disk buildup taken into account. In the model with the dust buildup taken into account, the midplane dust-to-gas ratio reaches its maximum value of 0.25 at around 0.25 Myr after the disk reached its maximum mass. The difference is caused by the dust evolution during the buildup stage, when the water ice-dominated dust present in the disk before the CO ice could condense had already grown and drifted inwards. What is more, planetesimal formation at the water snow line is active already at the buildup stage in this model, binding some pebbles and reducing the amount of dust available overall. A similar conclusion was reached by \citet{Charnoz2019}, who presented an analogous model including disk buildup and dead zone, and did not observe planetesimal formation at the CO snow line.

\begin{figure}
   \centering
   \includegraphics[width=\hsize]{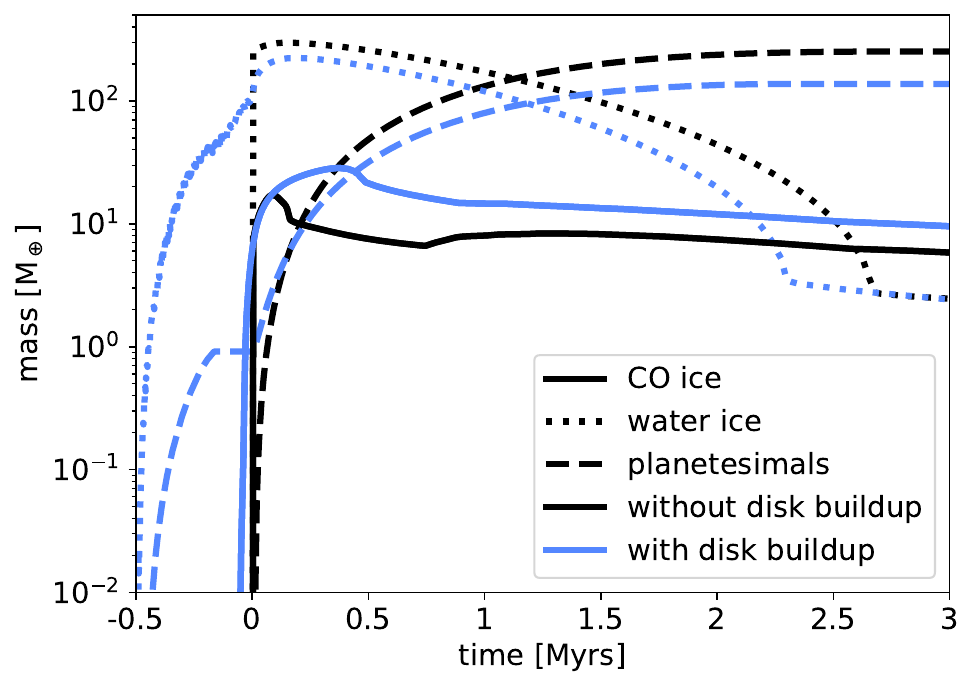}
      \caption{Time evolution of the total mass of CO ice, water ice, and planetesimals (as indicated by the different line styles) in the models with and without disk buildup (indicated with blue and black lines, respectively) for the set of models with the lower dust diffusivity.}
         \label{fig:massbudget_plts}
\end{figure}

\begin{figure*}
        \centering
        \includegraphics[width=0.85\textwidth]{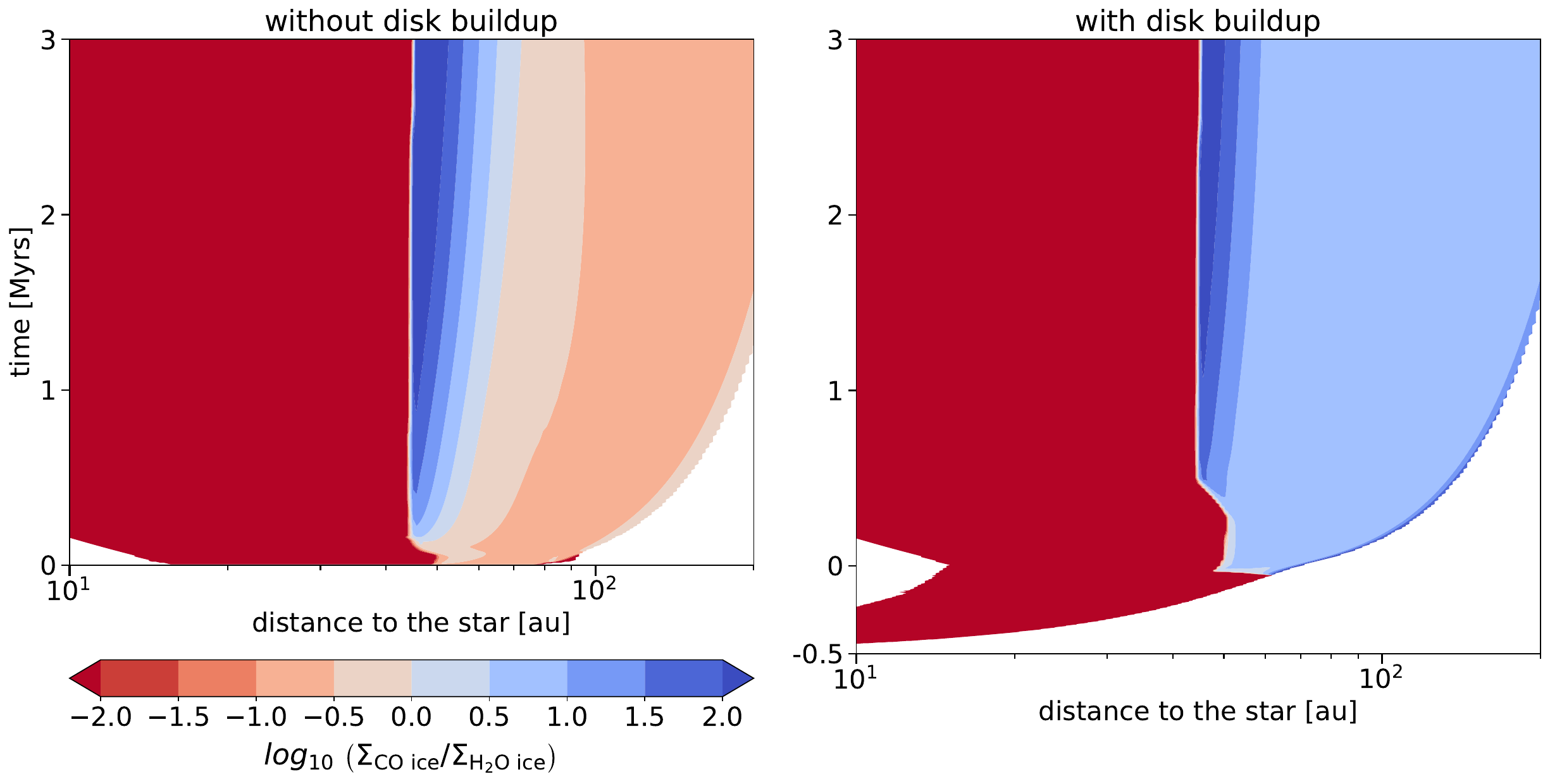}
        \caption{Evolution of the CO to H$_2$O ice ratio as a function of time and space in the models with lower dust diffusivity. The left panel corresponds to the model without disk buildup and the right panel to the model with disk buildup taken into account.}
        \label{fig:COtoH2O_plts}
\end{figure*}

Figure \ref{fig:massbudget_plts} shows the mass budget of CO ice, water ice, and planetesimals in the model assuming the lower dust diffusivity. The evolution of CO ice is similar to the fiducial model (see Fig.~\ref{fig:massbudget_fid}) as the CO ice is not bound in planetesimals. The evolution of water ice is quite different, as in the model with disk buildup, about 1 Earth mass of planetesimals already forms along the water snow line during the buildup stage. As in the fiducial model, there is less water ice at the time the model without buildup is initialized, but this time the difference propagates for over 2 Myrs. This is because now the water ice is not only lost to radial drift but also used for planetesimal formation. However, we limit the planetesimal formation rate such that only part of the available pebbles in the region that fulfill the planetesimal formation criteria are changed into planetesimals. Ultimately, there is about 100 Earth masses more planetesimals formed in the model without disk buildup, and the difference in water ice budget is erased after planetesimal formation stops at about 2.5~Myr of evolution. 

Figure~\ref{fig:COtoH2O_plts} shows that the effect of CO ice-dominated outer disk is even stronger for the models with low dust diffusivity as compared to the fiducial set. CO/H$_2$O ratios of several hundred, matching values observed in C/2016 R2, are obtained beyond the CO snow line, both in the model without and with disk buildup. The difference between the models with and without buildup is more pronounced than in the fiducial set as now both the CO ice augmentation and the water ice depletion play a role. In the model with the disk buildup included, the region outside of the CO snow line has CO/H$_2$O ratio of 5, while in the model without disk buildup, the CO/H$_2$O ratio drops to the fiducial value of 0.28 in the outer disk. However, as discussed above, no CO-rich  planetesimals are formed in our models.

\section{Discussion}\label{sect:discussion}

\subsection{Limitations of the model}

The results presented above are obtained using a one-dimensional model of protoplanetary disk evolution, which necessarily involves a number of simplifying assumptions. While this approach cannot capture non-axisymmetric structures, it allows us to isolate the fundamental effects of disk buildup and dust evolution on water and CO ice abundances. 

We need to stress, however, that the buildup model adopted here is rather simplistic. We assume that the initial molecular cloud is spherically symmetric, has a constant density, and rotates as a solid body. Furthermore, we assume that the angular momentum is conserved during the infall. The current observations and models of star-forming environments paint a more complex picture, with filamentary structures leading to heterogeneous accretion on disks \citep{Kuffmeier2017, Lee2021, Pineda2023, Bhandare2024}. It is not excluded that the infall itself leads to producing disk with substructure able to impact the dust evolution and planetesimal formation \citep{Kuznetsova2022, Huhn2025, Zhao2025}. 

In the model, we keep the luminosity of the central star constant. The evolution of the stellar luminosity would not have much impact on the location of the water snow line, which is located in the region dominated by viscous heating, but would have an effect on the location of the CO line, which is in the outer disk where the temperature is set by stellar irradiation \citep{Miley2021}. In the models presented by \citet{Price2021} that included the stellar luminosity evolution, the CO snow line moves inward with time. The inward movement of the CO line could lead to a lower efficiency of the cold-finger process with time, but should not significantly impact the results presented here.

Our model neglects the trapping of CO ice in amorphous water ice or clathrates found in laboratory experiments \citep{BarNun1985, Zhou2024, Ligterink2024, Pesciotta2024}. This effect can increase the effective binding energy of CO and partially couple its evolution to that of water \citep{Williams2025}. Including this process would likely reduce the degree of CO to water fractionation by allowing a fraction of CO to be removed together with drifting water-rich pebbles. However, because CO trapping requires the presence of water ice, its efficiency is expected to decrease in the models including infall, as early dust growth and drift progressively deplete the disk of water ice. Therefore, one can expect the qualitative trend of enhanced CO/H$_2$O ratios produced during disk buildup to be robust, although the absolute enrichment may be reduced.

We neglected chemical conversion of CO, which may be efficient in warm, UV-irradiated surface layers of the protoplanetary disk. However, chemical models show that CO conversion is expected to proceed slowly in the cold midplane regions considered here \citep{Eistrup2018, Bosman2018, VanClepper2022}.

\subsection{Planetesimal formation beyond the CO snow line}

CO-rich pebbles are formed at the CO snow line in both models with and without disk buildup. However, these pebbles are not participating in planetesimal formation, which only happens at the water snow line in the models assuming low dust diffusivity. Consequently, we do not produce planetesimals with a high CO/H$_2$O ratio that could be precursors of CO-rich comets such as C/2016 R2. Such bodies could potentially be produced if the infall was not homogeneous and triggered pressure traps in the disk \citep{Kuznetsova2022, Huhn2025, Zhao2025}. 

CO-rich bodies could also form as second-generation planetesimals. Our current model does not include the possibility of planetesimals growth to planets. If the planetesimals formed at the water snow line can form a massive planetary core that starts gas accretion and opens a gap in the disk, the outer edge of the planetary gap would act as a pressure bump that stops the pebble drift and induces planetesimal formation in the outer disk \citep{Eriksson2020, Shibaike2020, Lau2024}. Recently, \citet{Lau2025} showed that the combination of giant planet formation in the outer disk and disk dispersal by photoevaporation can lead to a late burst of planetesimal formation in the outer disk that will bind the leftover pebbles in planetesimals. In addition, photoevaporation alone, even in the absence of a planet-induced gap, may create pressure gradients capable of concentrating solids and promoting planetesimal formation \citep{Carrera2017, Ercolano2017, Ying2026}. These late-formed planetesimals would be spared the internal heating and processing by the short-lived radioactive isotopes, consistent with the properties of comets \citep{Prialnik1987}. Thus, the CO-rich pebbles could be turned into planetesimals if the full picture of planet formation and disk dispersal is taken into account, which is a motivation for future work on connecting disk evolution, planet formation, and chemistry.

\section{Conclusions}\label{sect:conclusions}

In this work, we have analyzed the prospect of forming planetesimals enriched in CO ice that could be the parent bodies of comets such as the comet C/2016 R2 (PanSTARRS). We found that pebbles with a very high CO/H$_2$O, matching the value of 300 observed in C/2016 R2, can be formed due to the cold finger effect outside of the CO snow line if the dust diffusivity is low. In the models including disk buildup, the whole region outside of the CO snow line becomes dominated by CO ice with CO/H$_2$O ratio of 3 to 5, depending on dust diffusivity. However, these CO-rich pebbles are not turned into planetesimals in our models. Planetesimals are only formed along the water snow line, as the dust-to-gas ratio is never high enough for planetesimal formation at the CO snow line. This suggests that CO-rich planetesimals may form preferentially as second-generation bodies, after CO enrichment has occurred and if solid concentrations are locally enhanced. Such enhancement could occur in pressure bumps, planet-induced gaps, or regions affected by photoevaporative gas removal, but these mechanisms were not modeled here.

Comparing the standard model starting with a fully-fledged protoplanetary disk and initialized with a constant dust-to-gas ratio to models including disk buildup, we found significant differences regarding the CO/H$_2$O ratio of pebbles in the outer disk. Dust evolution during these earliest stages of disk buildup leads to a lower water ice abundance, smaller grain size, and augmented CO condensation compared to the prediction of a model not including disk buildup. Similar behavior should be expected for other volatiles. Thus, models concerning the chemical evolution of the disk should consider the buildup stage in the efforts of constraining planetary formation from the observed atmospheric molecular ratios such as C/O \citep[see, e.g.,][]{Oberg2011, Booth2017, Krijt2020, Schneider2021, Houge2025, Guzman2026}.

\begin{acknowledgements}
JD thanks the anonymous referee for constructive comments that helped to improve this paper. JD was funded by the European Union under the European Union’s Horizon Europe Research \& Innovation Programme 101040037 (PLANETOIDS). Views and opinions expressed are however those of the author only and do not necessarily reflect those of the European Union or the European Research Council. Neither the European Union nor the granting authority can be held responsible for them.
 
\end{acknowledgements}

\bibliographystyle{aa} 
\bibliography{COvswater}

\end{document}